\documentclass[twoside]{ilcws08}
\usepackage[latin1]{inputenc}
\usepackage[dvips]{graphicx,epsfig,color}
\usepackage{wrapfig,rotating}
\usepackage{amssymb,amsmath,array}

\begin{document}
\title{Experience with the AHCAL Calibration System in the Test Beam}
\author{G. Eigen, T.Buanes
\thanks{We thank the DESY AHCAL group. This work was supported by the Norwegian Research Council.}
\vspace{.3cm}\\
Representing the CALICE collaboration \\
Department of Physics, Allegaten 55 \\
Bergen, Norway \\ 
\vspace{.1cm}}

\maketitle

\begin{abstract}
We present herein our experience with the calibration system in the CALICE AHCAL prototype in the test beam and discuss characterizations of the SiPM response curves. 
\end{abstract}

\section{Introduction}

SiPMs are excellent photosensors for low-intensity light detection. With increasing light intensities, however, their response becomes non linear requiring monitoring. In the CALICE AHCAL prototype we installed an LED/PIN-diode-based monitoring system to measure the SiPM gain, monitor the SiPM response for fixed light intensities and record the full SiPM response function when necessary. We also  record the reverse bias voltage of each SiPM and the temperature measured by five sensors in each layer with a slow-control system, since the SiPM response is very sensitive to changes in these parameters. We studied the SiPM response function in test beam calibration runs. If we find an analytic function that parameterizes the SiPM response, the monitoring system might be simplified. The exact shape would be measured once on the test bench before installation.

\section{Calibration of an AHCAL Cell}

The raw energy in a cell measured in units of ADC bins, $Q^{meas}_{cell}[ADC]$, is converted into units of MIPs via a calibration factor $C^{MIP}_{cell} [ADC]$ 

\begin{equation}
E^{meas}_{cell}[MIP] =\frac{Q^{meas}_{cell}[ADC]}{C^{MIP}_{cell}[ADC]} \cdot f^{-1}_{sat}(Q^{meas}_{cell}[pixel]),
\end{equation}
\noindent
where $f_{sat}(Q^{meas}_{cell}[pixel])$ is the non-linear response parameterized as a function of pixels. It includes an intercalibration factor matching the pixel and MIP calibration scales \cite{minical}. We measure the SiPM gain with low-intensity LED light, the position of the MIP peak with muons and the non-linearity by varying the LED light intensity. Gain and light yield vary with temperature (reverse bias voltage) as $1/G \cdot dG/dT =-1.7\%/K~(1/G \cdot dG/dV=2.5\% /0.1V$) and $1/Q \cdot dQ/dT =-4.5\% /K~(1/Q \cdot dQ/dV=7.0\% /0.1V$), respectively.

\begin{figure}
\includegraphics[width=0.48\textwidth]{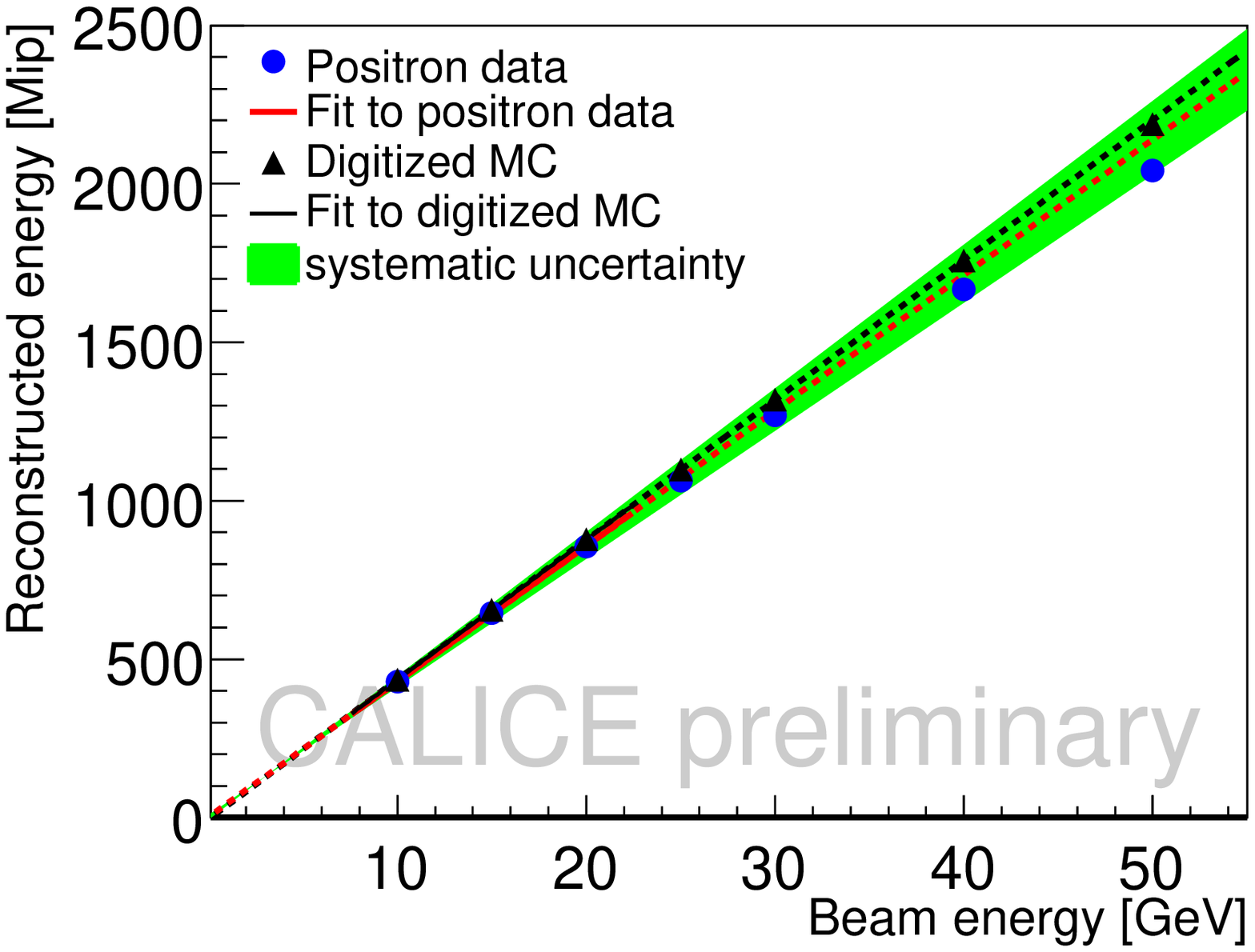}
\includegraphics[width=0.48\textwidth]{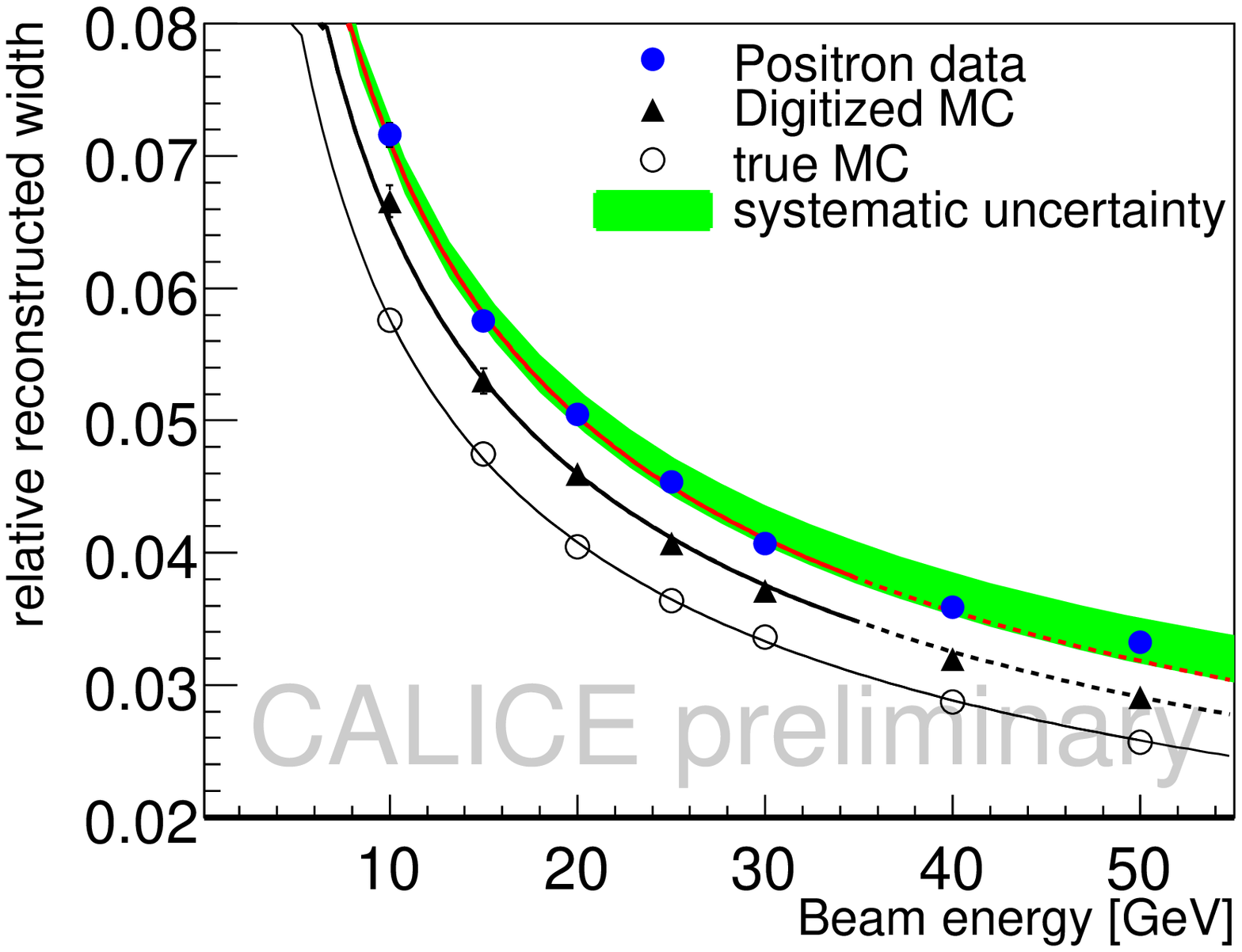}
\caption{Reconstructed energy (left) and energy resolution (right) for $e^+$ data (dots), bare simulations (open circles) and simulations including detector effects (triangles). Lines show fits and shaded regions systematic uncertainties.}\label{fig:energy}
\end{figure}

The standard procedure consists of adjusting $G$ and $Q$ for temperature changes. Here, corrections are instantaneous but non-local, since $T$ is measured frequently at five positions per layer during a run. Since the gain of each cell is measured several times a day, we could also correct $Q$ for gain changes by $Q =Q_0 + dQ/dG \cdot \Delta G$ \cite{feege}. This procedure is local but not instantaneous. Figure~\ref{fig:energy} shows energy and energy resolution measurements of 10-- 50~GeV positrons in comparison to simulations \cite{enote}. The data is corrected for saturation and temperature effects using cell-wise gain and MIP calibration constants and layer-wise average temperature correction factors. The reconstructed energy is linear up to 30~GeV. At 50~GeV deviations from linearity increase to $\sim 5\%$ suggesting the need for a refined analysis procedure. The energy resolution fits the standard form containing a stochastic term, a constant and a noise term. The bare simulations demonstrate that detector effects are important. Simulations that include detector effects, however, are still too optimistic and require refinements. 


\section{Study of the SiPM Response Function}

We extract SiPM and PIN diode raw data from LCIO files, perform pedestal subtraction with VCalib runs, apply gain corrections, and use intercalibration constants. For each Vcalib value we perform Gaussian fits to the SiPM and PIN diode response to determine mean values and their errors. We plot the PIN-diode response versus the SiPM response after rescaling the PIN-diode values to start at a common origin with a slope of one. So far we analyzed four calibration runs from the CERN test beam in 2006 and 2007 \cite{buanes}.

\begin{wrapfigure}{r}{0.5\columnwidth}
\centerline{\includegraphics[width=0.46\columnwidth]{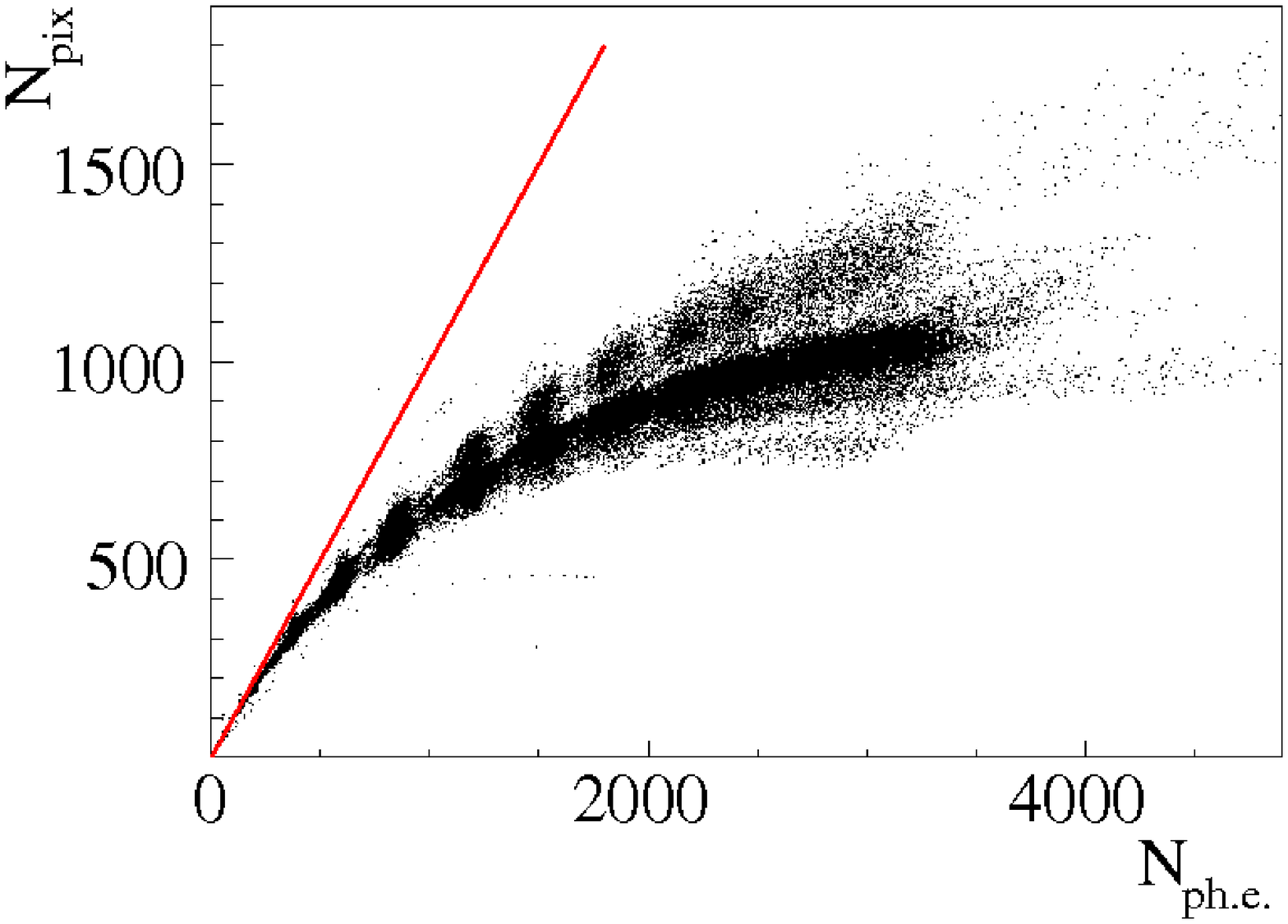}}
\centerline{\includegraphics[width=0.45\columnwidth]{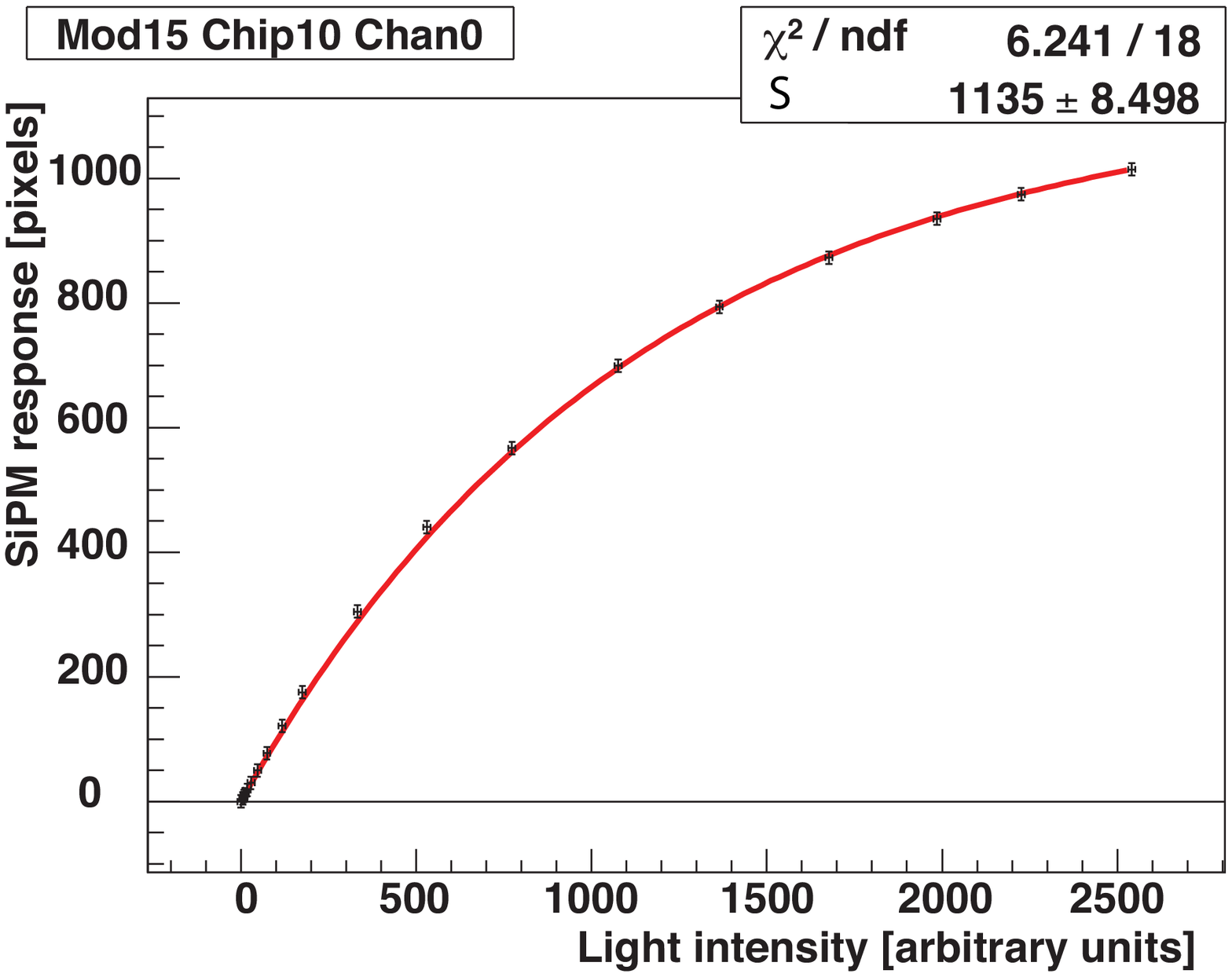}}
\centerline{\includegraphics[width=0.45\columnwidth]{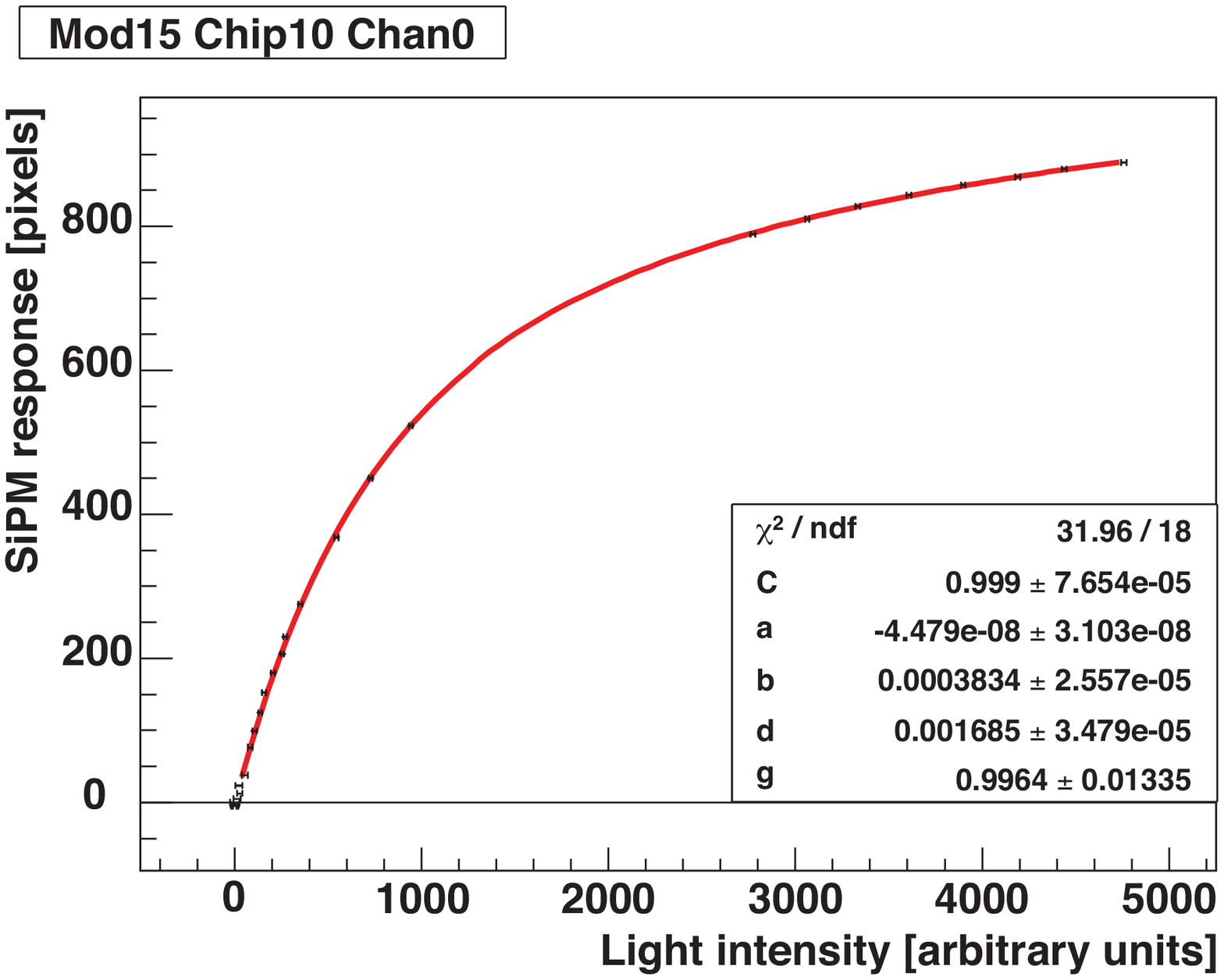}}
\caption{SiPM response curves from ITEP (top);  measurements of a SiPM response curve with superimposed fit from ITEP (middle) and test beam (bottom).}\label{fig:curves}
\end{wrapfigure}

At ITEP, the response of all 7608 SiPMs was measured prior to installation into the AHCAL prototype using calibrated LED light shone directly onto the SiPM. Figure~\ref{fig:curves} (top) shows the ITEP measurements after scaling raw data to start at a common origin with slope one. The curves fit to $f_I (x)=S(1-exp(-ax))$, where the saturation $S$ and $a$ are free parameters. In the test beam the SiPM response is measured with high-gain and low-gain preamplifier settings. Since the curves do not fit to $f_I (x)$, we constructed the function

\begin{equation}
f_B(x)= \frac{1}{g(x)}\bigl[\frac{(C-1)^2}{a-(b+d)(C-1)}\cdot \frac{\exp(-bx)+\exp(-dx)}{C-\exp(ax)}-\frac{2(C-1)}{a-(b+d)(C-1)} \bigr],
\end{equation}
where $C, a, b, d$  (with $b<d)$ and $g$ are free parameters in the fit. The latter is a scale factor that 
accounts for mismatch between high-gain and low-gain regions. With simpler forms we obtained fewer successful fits.

Figures~\ref{fig:curves}(middle, bottom) show response curves of a typical SiPM for ITEP and test beam data, respectively. Fitting all available SiPMs in the four calibration runs, we get 50--60$\%$ successful fits. In the 31-07-07 run $60\%$ of all fits were successful, $30\%$ failed due to a small confidence level and $10\%$ failed due to malfunctioning of the SiPM or PIN diode. For successful fits, parameter $b$ ($d$) peaks at $\sim 0.00025$~$(0.0016)$ with FWHM of $ \sim 0.0002~(0.0035)$. Apart from a few outliers at higher values, a spike near zero is visible. Parameter $a$ peaks near $3\times 10^{-7}$ with FWHM of $ \sim 6 \times 10^{-7}$, while parameter $C~(g)$ lies around $0.998~(0.99)$ with FWHM of $\sim 0.001~(0.08)$. 

\begin{figure}
\includegraphics[width=0.49\textwidth]{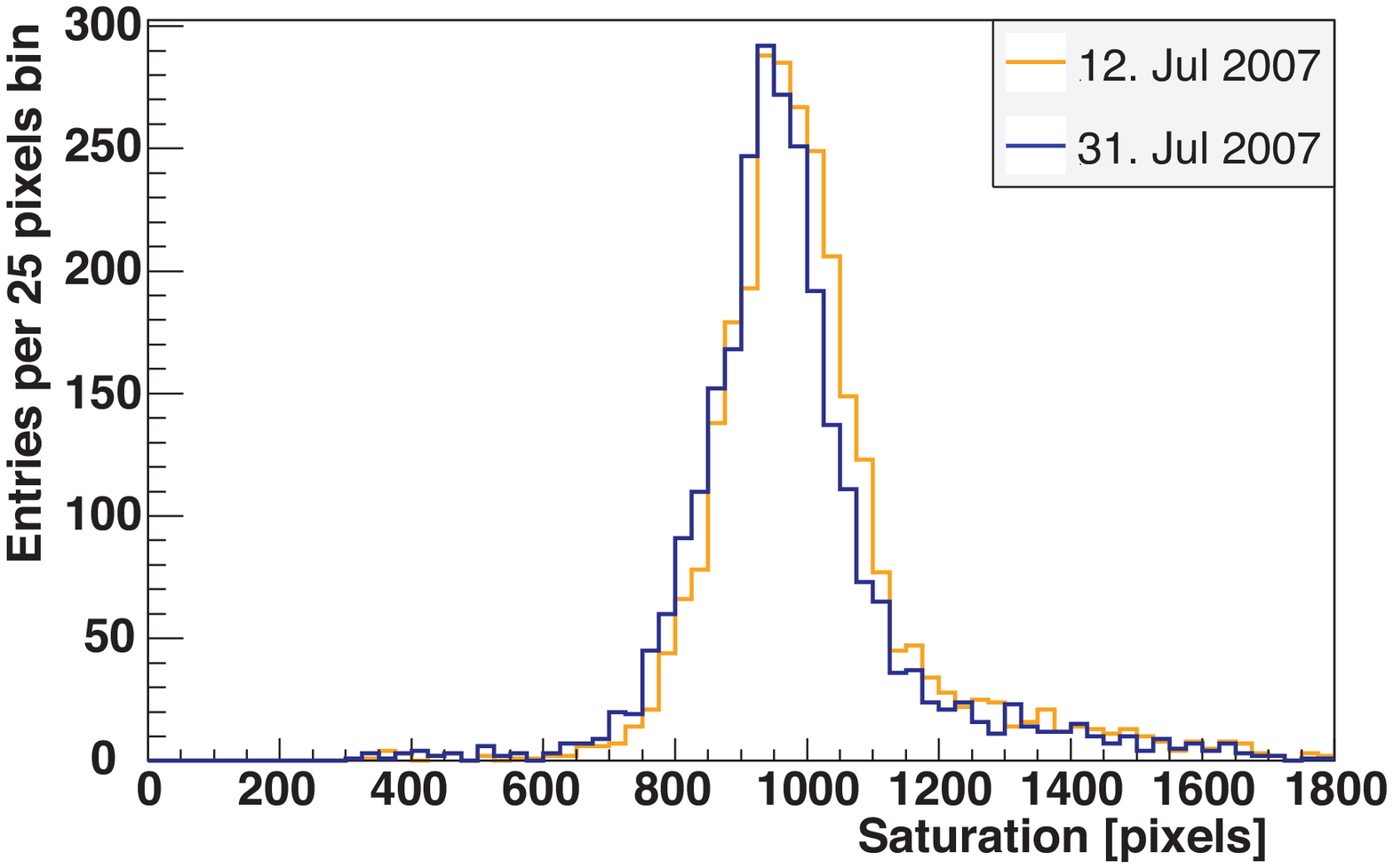}
\includegraphics[width=0.49\textwidth]{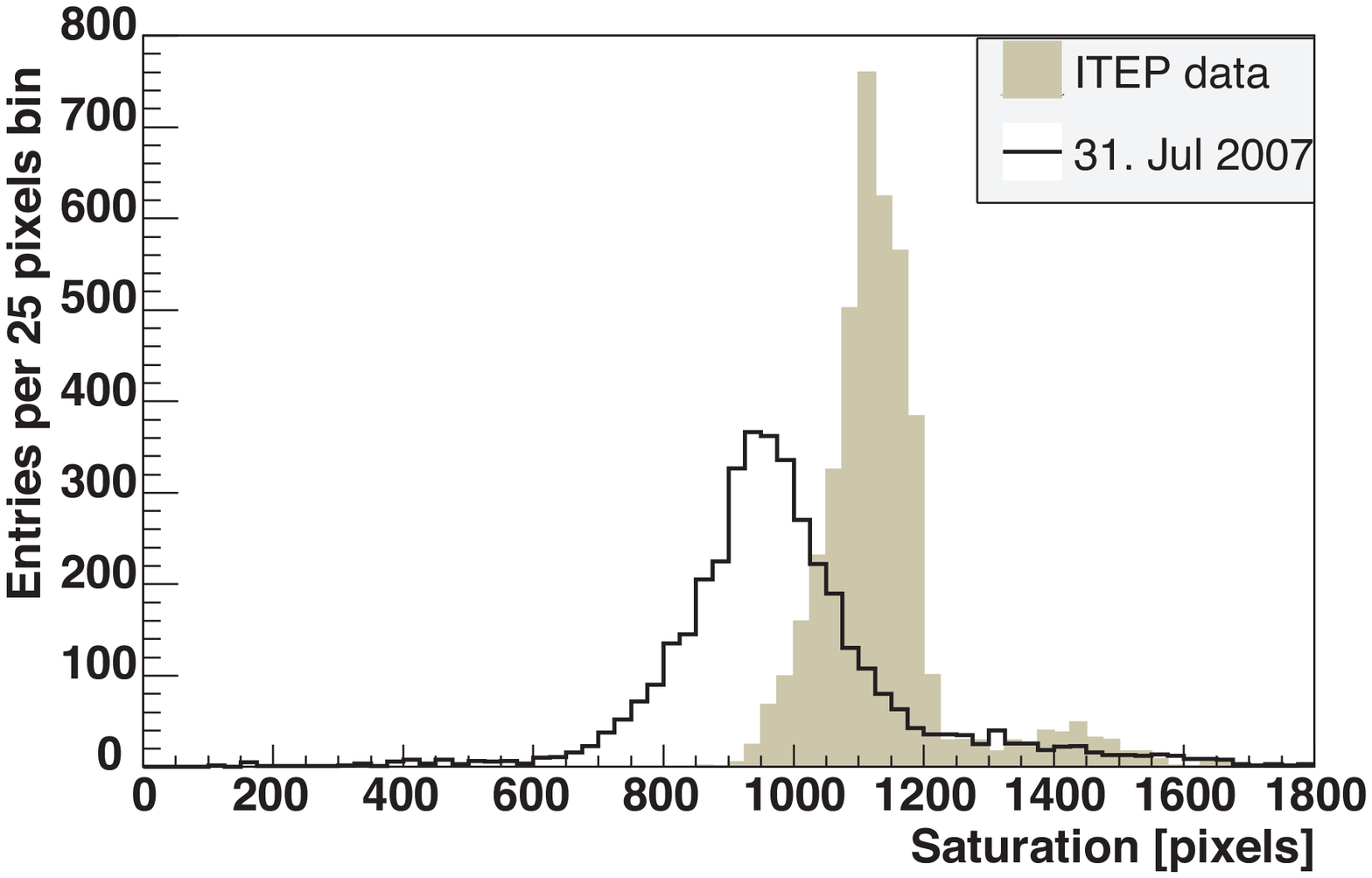}
\caption{Saturation values for two test beam calibration runs (left) and a comparison of the saturation measured in the test beam and at ITEP (right).}\label{fig:saturation}
\end{figure}

Figure~\ref{fig:saturation}(left) shows saturation values for 12-07-07 and 31-07-07 calibration runs. For the later run, the saturation peaks near $930$ pixels and has a width of FWHM$ \sim 180$ pixels, while for the earlier run the peak is shifted upward by $\sim 25$ pixels. This is consistent with expected temperature variations that are not corrected for. In addition, we see a long tail up to $1800$ pixels in both runs that is also visible in ITEP data. This results from a particular batch of SiPMs with a lower internal resistor causing multiple pixel excitations during illumination.  A comparison of saturation values of SiPMs in modules 3--15 for four runs taken in 2006 and 2006 shows no degradation thus confirming
stable operation in the test beam. 

Figure~\ref{fig:saturation}(right) shows a comparison of saturation values measured in the 31-07-07 run and at ITEP. The ITEP data  peak around $1100$ pixels with a FWHM of $\sim 50$ pixels indicating that nearly all 1156 pixels in the SiPMs are triggered. Here, the SiPMs were illuminated directly by LEDs, while the light transport in the test beam is rather complex. The SiPM records the light from a 1~mm thick wavelength-shifting (WLS) fiber coupled via an air gap. Though a 0.2~mm air gap is sufficient for a full illumination of all pixels, losses may occur due to imperfect alignment of the fiber and SiPM.  Because of the large discrepancy between ITEP and test beam data, the non-linearity corrections are presently based on the ITEP saturation measurements with an additional scaling by the ratio of saturation values measured for each SiPM in the test beam and at ITEP.

\section{Conclusion and Outlook}

We found an analytical function that successfully parameterizes about $60\%$ of the SiPM response curves in the CALICE AHCAL prototype. In order to improve this close to $90\%$ we need to investigate causes for poor fits and test other analytical functions. The present studies show that the SiPM operation in the CERN test beam is stable. We observe a discrepancy of saturation values measured at ITEP and in the test beam of about $15\%$, which may be caused by a misalignment of the WLS fiber and the SiPM. We need to verify that we can model temperature and voltage changes in the SiPM response curves and we need to extend our studies to include 2008 test beam data at Fermilab.

At the present level of understanding a full monitoring system is necessary that allows us to measure the full SiPM response any time. The system, however, may be simplified. Two options are under discussion. The first option is based on the present monitoring system but foresees long clear fibers, each illuminating one row of tiles rather than individual tiles. This would reduce the number of LEDs but it may not achieve sufficient light intensities in all tiles (see J. Zalesak's talk~\cite{zalesak}). The second option consists of embedding one LED per tile eliminating fibers but requiring a huge number of LED's. System tests started to optimize LED positions, check the homogeneity of the response and test different LED types. The light calibration is compared to the response of a radioactive source. The system will be temperature controlled. First tests show no cross-talk, but  optimization for dynamic range and LED uniformity is needed.

\end{document}